\documentclass[a4paper, superscriptaddress, amsfonts, amssymb, amsmath, reprint, showkeys, nofootinbib, twoside]{revtex4-1}

\usepackage{graphicx}
\usepackage{dcolumn}
\usepackage{bm}
\usepackage[T1]{fontenc}
\usepackage{siunitx}
    \usepackage{braket}
    
    \usepackage{color}
    \usepackage{xcolor}
    \usepackage{multirow}
    \usepackage{amsmath}

    \usepackage{hhline}
    \usepackage{threeparttable}
    \usepackage{hyperref}
    
    \def\U#1{{\rm #1}} 
    \def\u#1{_{\rm #1}}

    \newcommand{\od}[2]{\frac{\mathrm{d} #1}{\mathrm{d} #2}}

\begin{document}


\title{Channel-selective frequency up-conversion \\for frequency-multiplexed quantum network}

\author{Shoichi Murakami}
 \email{smurakami@qi.mp.es.osaka-u.ac.jp}
 \affiliation{ Graduate School of Engineering Science, Osaka University, Osaka 560-8531, Japan}%
  \affiliation{ Center for Quantum Information and Quantum Biology, Osaka University, Osaka 560-0043, Japan}%

\author{Shunsuke Hiraoka}
 \affiliation{ Graduate School of Engineering Science, Osaka University, Osaka 560-8531, Japan}%
 
\author{Toshiki Kobayashi}%
  \affiliation{ Graduate School of Engineering Science, Osaka University, Osaka 560-8531, Japan}%
  \affiliation{ Center for Quantum Information and Quantum Biology, Osaka University, Osaka 560-0043, Japan}%

\author{\\Takashi Yamamoto}%
  \affiliation{ Graduate School of Engineering Science, Osaka University, Osaka 560-8531, Japan}%
  \affiliation{ Center for Quantum Information and Quantum Biology, Osaka University, Osaka 560-0043, Japan}%

\author{Rikizo Ikuta}%
  \affiliation{ Graduate School of Engineering Science, Osaka University, Osaka 560-8531, Japan}%
  \affiliation{ Center for Quantum Information and Quantum Biology, Osaka University, Osaka 560-0043, Japan}%

\date{\today}

\begin{abstract}
We demonstrate channel-selective frequency up-conversion from telecom wavelengths around \SI{1540}{nm} for optical fiber communication to visible wavelengths around \SI{780}{nm}, based on second-order optical nonlinearity in a cavity of the converted modes. 
In our experiment, we selectively convert a light from any frequency mode within frequency-multiplexed telecom signals to a desired output mode, determined by the cavity resonances. 
Based on the experimental results of the frequency up-conversion, we derive the signal-to-noise ratio of the process at the single-photon level, and discuss its applicability to channel-selective quantum frequency conversion~(CS-QFC) in the context of frequency-multiplexed quantum networks. 
Finally, we describe specific use cases of the CS-QFC, which show its utility as a reconfigurable switching element in frequency-multiplexed networks, 
particularly for selectively performing Bell-state measurements between two photons originating from different frequencies. 
\end{abstract}

\maketitle
\section{Introduction}
\label{intro}
Quantum frequency conversion~(QFC) of photons~\cite{kumar1990quantum} is a promising approach to link various kinds of quantum systems, such as neutral atoms~\cite{ikuta2018polarization}, ions~\cite{bock2018high}, NV centers~\cite{tchebotareva2019entanglement}, SiV centers~\cite{bersin2024telecom}, and so on, because each of the quantum systems interacts only with a photon at a frequency specific to its physical characteristics. 
With the aim of realizing an optical-fiber-based quantum internet~\cite{kimble2008quantum,wehner2018quantum,Azuma2023} over long distances, including quantum repeaters with quantum memories~\cite{duan2001long}, QFC is essential for bridging the frequency gap between photons emitted from quantum systems and those in the telecommunication band.
So far, extensive experimental research on QFC has focused particularly on connecting visible and near-infrared wavelengths to the telecom band~\cite{tanzilli2005photonic, ikuta2011wide}.
Recently, successful entanglement generation between remote quantum systems via entanglement swapping based on the Bell-state measurement~(BSM) between telecom photons after QFC has been reported~\cite{Yu2020-va,van2022entangling,stolk2024metropolitan,Liu2024-oz,strobel2024quantum,luo2025entangling}. 
In these experiments, each emitted photon is converted to the same telecom wavelength via frequency down-conversion using the second-order nonlinear optical medium 
with a strong pump light at a frequency corresponding to the difference between the input and output frequencies. 

For building efficient and flexible quantum communication systems, it is crucial to incorporate frequency-division multiplexing techniques into quantum networks involving entanglement swapping.
Many experiments related to the generation and direct distribution of frequency-multiplexed entangled photon pairs over multiple frequency channels have been demonstrated~\cite{wengerowsky2018entanglement,joshi2020trusted,qi202115,fujimoto2022entanglement,Chang2025}. Beyond these experiments, 
frequency-multiplexed two-photon interference aimed at BSM has also been reported~\cite{Ichihara2023,Huang2025,Yan2025}, although within the classically observable visibility regime. 
In these experiments, interference has been observed between photons originating from the same frequency channels using passively designed measurement systems. 
However, inspired by classical network systems, it is expected that large-scale frequency-multiplexed quantum networks will require not only passive architectures but also reconfigurable functionalities~\cite{lingaraju2021adaptive,alshowkan2021reconfigurable,Sakuma2024,Liu2024,Shapourian2025}, 
which will enable the network to be flexibly and efficiently expanded to support massive multi-user configurations and additional nodes. 

\begin{figure}[t]
    \centering
    \includegraphics[width = \linewidth]{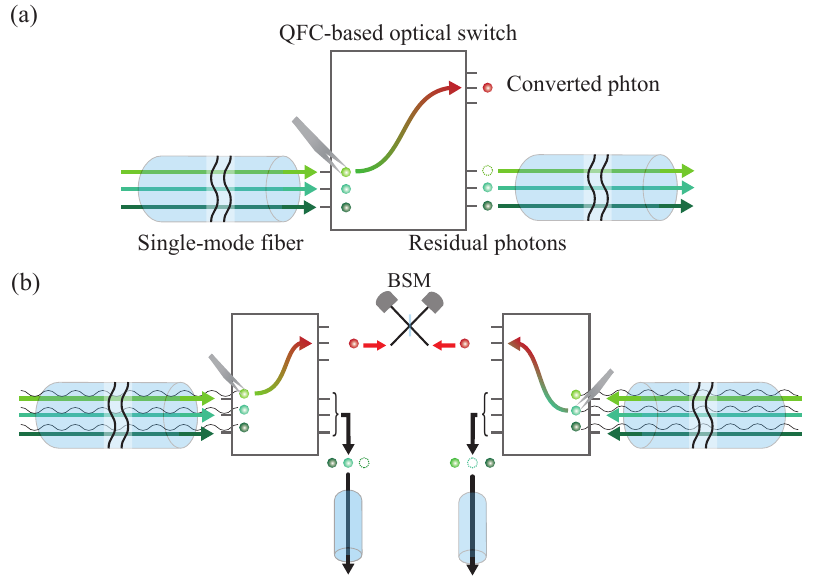}
    \caption{(a) 
    Concept of the QFC-based optical switch. 
    This device extracts only a photon in a specific frequency bin from multiplexed frequency channels transmitted through an optical fiber, 
    while leaving the remaining frequency modes unaffected and passing them on to the subsequent transmission channel.
    (b) Concept of the channel-selective BSM using the QFC-based optical switch in a frequency-multiplexed quantum network. }
    \label{fig:concept} 
\end{figure}
In the frequency-multiplexed reconfigurable quantum networks, the two distributed photons involved in BSM for entanglement swapping may not always be in the same frequency channel~\cite{qi2025multiuser}. 
Therefore, similar to reconfigurable optical add/drop multiplexers~(ROADMs)~\cite{hattori1999plc,ford1999wavelength,ertel2006design} widely used in the current fiber-based telecom network technologies, a QFC-based optical switch, as illustrated in Fig.~\ref{fig:concept}~(a), that converts one of the frequency-multiplexed telecom photons to a photon at a different output wavelength is desirable. 
The QFC-based optical switch should possess the key abilities 
(i) to address a single frequency mode within the frequency-multiplexed channels; 
(ii) to convert a photon in the mode into a target frequency to enable a BSM;
(iii) to preserve photonic states in all other frequency modes to avoid disturbing ongoing quantum communication in other channels; and 
(iv) to dynamically reconfigure the switching configuration on demand in response to network requirements or user demand. 
The conceptual figure of the BSM system based on the QFC-based optical switch for a frequency-multiplexed network is shown in Fig.~\ref{fig:concept}~(b). 
By QFC, any two photons from arbitrary frequency channels can be extracted and measured in the same frequency mode for BSM, without disturbing photons in other frequency channels. 
Beyond the BSM system, the QFC-based optical switch offers a variety of applications in frequency-multiplexed quantum networks, including those involving matter-based quantum systems.
Owing to the broad scope of these applications, further details are provided in the Discussion section.

For this purpose, we experimentally demonstrate such a channel-selective frequency up-conversion using a laser light from the telecom band around \SI{1540}{nm} to \SI{780}{nm} via sum frequency generation~(SFG) in a periodically poled lithium niobate~(PPLN) waveguide with a cavity structure around \SI{780}{nm}. 
Unlike channel-selective QFCs~(CS-QFCs) demonstrated in  Refs.~\cite{wang2021proposal,arizono20241xn} without a cavity structure, 
we have successfully demonstrated fully selective frequency up-conversion of light, owing to the cavity-assisted conversion setup, proposed as optical frequency tweezers~\cite{ikuta2022optical}. 
This enables arbitrary selection of the single frequency mode from a frequency-multiplexed input and of the converted output mode among the cavity-resonant modes by tuning the pump frequency.
Based on the experimental results, we estimate the signal-to-noise ratio~(SNR) of the frequency up-conversion process, expected for input multiplexed signals at the single-photon level, and discuss applicability of our device to CS-QFC in frequency-multiplexed quantum networks. 
We present several use cases of the CS-QFC beyond channel-selective BSMs between photons from different frequency channels.

\begin{figure}[t!]
    \centering
    \includegraphics[width = \linewidth]{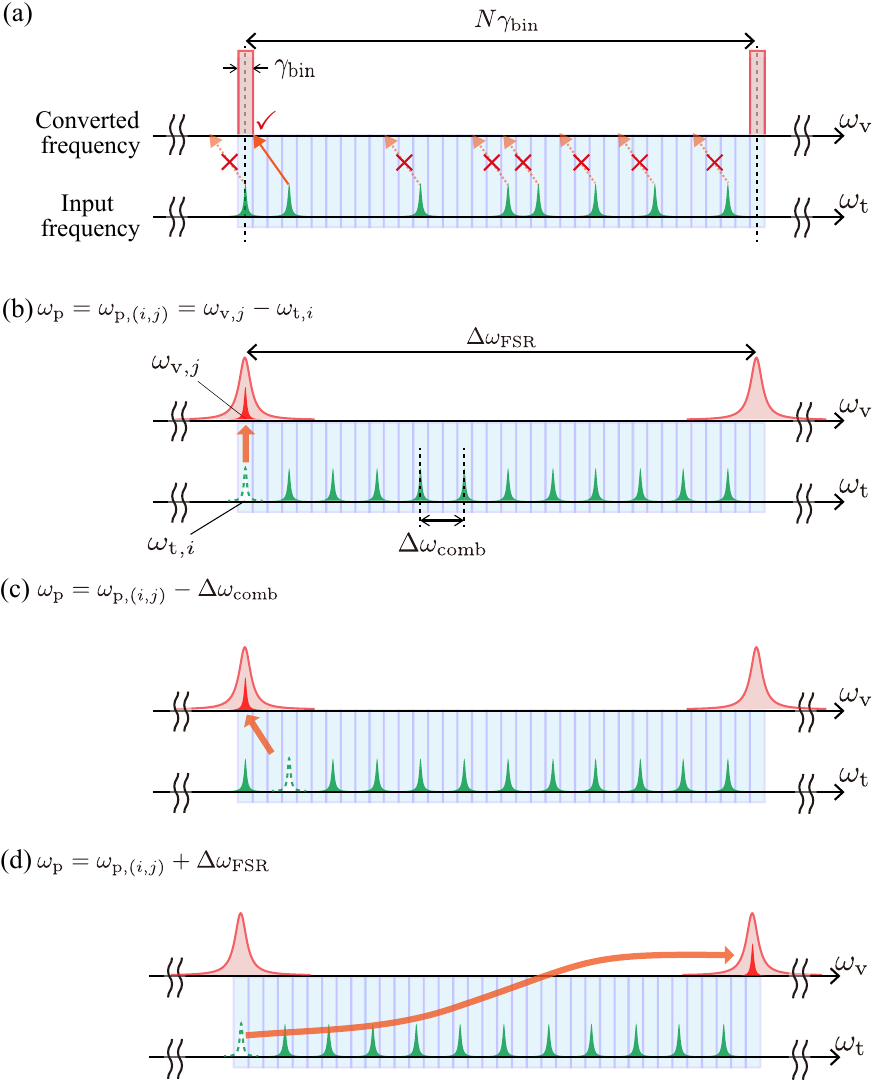}
    \caption{
    (a)~Design concept of CS-QFC.
    The red rectangular areas each with the width of $\gamma\u{bin}$ 
    represent the destinations accessible through the conversion process.
    Among the input photons distributed over multiple frequency bins~(green wave packets), only a photon in a specific frequency bin is converted in a single tweezing process, and the bin can be selectively chosen by tuning the pump frequency. Details are given in the main text. 
    (b-d)~Implementation of CS-QFC based on a SFG with a cavity structure   around the converted frequencies.
    These illustrate the tweezing operation of only the photons in a specific tooth of a frequency comb input with spacing $\Delta\omega\u{comb}$, by selecting the pump frequency. 
    (b) When the pump frequency $\omega\u{p}$ is $\omega_{\mathrm{p},(i,j)}(=\omega_{\mathrm{v},j}-\omega_{\mathrm{t},i})$, 
    only the photon centered at $\omega_{\mathrm{t},i}$ is converted.
    (c) For $\omega\u{p}=\omega_{\mathrm{p},(i,j)} - \Delta\omega_{\mathrm{comb}}$, only the photon at $\omega_{\mathrm{t},i}+\Delta\omega\u{comb}$ is converted. 
    (d) For $\omega\u{p}=\omega_{\mathrm{p},(i,j)}+\Delta\omega\u{FSR}$, 
    the photon at $\omega_{\mathrm{t},i}$ is converted to the neighboring bin   centered at $\omega_{\mathrm{v},j}+\Delta\omega\u{FSR}$. 
    }
    \label{fig:theory_tweezer} 
\end{figure}

\section{Design concept and implementation}\label{sec:theory}
\subsection{Design concept}\label{subsec:design_concept}
In the scenarios involving the optical frequency tweezers through frequency up-conversion considered in this paper, we use a single-frequency pump light to selectively tweeze an input photon from a target frequency bin within multiplexed frequency modes to a desired output frequency bin in each individual process. A conceptual illustration is shown in Fig.~\ref{fig:theory_tweezer}~(a).
We treat the accessible output frequencies for conversion as discrete bins, each with a finite width of $\gamma\u{bin}$. These bins are spaced apart by $N\gamma\u{bin}$,
where $N\gg 1$ is an integer, meaning that only specific frequency bins are allowed as conversion destinations.
On the input side, accessible frequency bins with a width of $\gamma\u{bin}$ are continuously arranged, allowing photons to occupy any number of frequency bins within $N\gamma\u{bin}$ simultaneously, with a maximal occupation number of $N$. 

We denote the pump frequency that maps the $i$-th input frequency bin centered at $\omega_{\U{in},i}$ to the $j$-th accessible output frequency bin 
centered at $\omega_{\U{out},j}$ by 
$\omega_{\U{p},(i,j)}(=\omega_{\U{out},j} - \omega_{\U{in},i})$. 
In this process, if a photon is in the $i$-th input frequency bin, 
it is converted to the $j$-th output frequency bin. 
In contrast, 
photons in input frequency bins labeled $i\pm n\gamma\u{bin}$ 
for $n=1,\ldots, N-1$ are never converted. 
Using the pump light at $\omega_{\U{p},(i',j)}$, 
only another photon in the $i'$-th input frequency bin can be converted 
without altering the output frequency bin. 
Similarly, if we use the pump light at $\omega_{\U{p},(i,j')}$, 
only the output frequency bin is changed, 
while the target input bin remains the same. 
We refer to this operation as the optical frequency tweezers~\cite{ikuta2022optical}, 
which enables selective frequency conversion.

It should be noted that the tweezing operation applies the same transformation to every $N\gamma\u{bin}$ input frequency bins. 
In the conversion process from the $i$-th to the $j$-th bin, the photons in the $(i+mN\gamma\u{bin})$-th input bins are mapped simultaneously to the $(j + m)$-th output bins for all integer $m$. 
Consequently, even if the frequency range of the multiplexed signal is extended, the maximum achievable multiplexing degree is remains limited to $N$ when attempting to tweeze a photon within a single frequency bin of a multiplexed signal.

\subsection{Implementation}
In order to realize this concept, 
we use the sum frequency generation~(SFG) based on second-order optical nonlinear interaction inside a cavity resonant at the converted frequency. 
As a first step, we describe the theory of frequency up-conversion 
from the signal mode $\hat{a}\u{t}$ at a single angular frequency $\omega\u{t}$ 
to the converted mode $\hat{a}\u{v}$ at angular frequency $\omega\u{v}$ based on the cavity-enhanced SFG process described above. 
The pump frequency for SFG is $\omega\u{p}(= \omega\u{v}-\omega\u{t})$. 
As introduced in Refs.~\cite{ikuta2022optical,murakami2025low,collett1984squeezing}, 
we describe the coupling constant between the inside and outside of the cavity by $\sqrt{\gamma\u{r}}$ which is related to the reflectance of the mirror. 
We denote the coupling constant of the SFG process by $\xi=|\xi|e^{i\phi}$, 
which is proportional to the complex amplitude of the pump light. 
In the model, the time evolution of the mode $\hat{a}\u{c}$ of the converted light 
inside the cavity, in a frame rotating at the resonant frequency of $\omega\u{c}$, 
is written as 
\begin{equation}\label{eq:diffeq_a_c}
    \od{\hat{a}\u{c}}{t}
=i\Delta \hat{a}\u{c} - \frac{\gamma_\mathrm{all}+|\xi|^2}{2}\hat{a}\u{c}+\xi \hat{a}_{\mathrm{t},IN} + \sqrt{\gamma_\mathrm{r}}\hat{a}_{\mathrm{v},IN},
\end{equation}
where $\Delta = \omega\u{v} - \omega\u{c}$ is the detuning of the converted mode,  $\gamma\u{all} = \gamma\u{r}+\gamma\u{int}$ is the total loss determined by $\gamma\u{r}$ and the internal loss $\gamma\u{int}$ of the cold cavity, and $\hat{a}_{\mathrm{t(v)},IN}$ is the annihilation operator denoting the input mode to the cavity.
Using the input-output relations 
$\hat{a}_{\mathrm{t},IN} - \hat{a}_{\mathrm{t},OUT} = \xi^*\hat{a}\u{c}$ and $\hat{a}_{\mathrm{v},IN} + \hat{a}_{\mathrm{v},OUT} =\sqrt{\gamma\u{r}}\hat{a}\u{c}$, the complex amplitudes of the converted and remaining modes are described by 
\begin{align}
\hat{a}_{\U{v},OUT}
&= 
\frac{e^{i\phi}\sqrt{\tilde{\gamma\u{r}}\tilde{C}}\hat{a}_{\U{t},IN}+\left[\tilde{\gamma}\u{r}-\frac{1}{2}(1+\tilde{C})+i\tilde{\Delta}\right]\hat{a}_{\U{v},IN}}{\frac{1}{2}(1+\tilde{C})-i\tilde{\Delta}},
\label{eq:av}\\
\hat{a}_{\U{t},OUT}
&= \frac{\left[\frac{1}{2}(1-\tilde{C})-i\tilde{\Delta}\right]\hat{a}_{\U{t},IN} + e^{-i\phi}\sqrt{\tilde{\gamma\u{r}}\tilde{C}}\hat{a}_{\U{v},IN}}{\frac{1}{2}(1+\tilde{C})-i\tilde{\Delta}}. 
\label{eq:at}
\end{align}
where $\tilde{\gamma}\u{r}=\gamma\u{r}/\gamma\u{all}$, $\tilde{C} = |\xi|^2/\gamma\u{all}$, $\tilde{\Delta} = \Delta/\gamma\u{all}$, and $\hat{a}_{\mathrm{t(v)},OUT}$ is the annihilation operator denoting the output mode from the cavity.
Assuming $a_{\U{v},IN}=0$ for frequency up-conversion from the signal mode $\hat{a}_{\mathrm{t},IN}$ to the converted mode $\hat{a}_{\mathrm{v},OUT}$ considered in this paper, Eq.~(\ref{eq:av}) is written by 
\begin{align}
\hat{a}_{\U{v},OUT}
&= 
\frac{e^{i\phi}\sqrt{\tilde{\gamma\u{r}}\tilde{C}}\hat{a}_{\U{t},IN}}{\frac{1}{2}(1+\tilde{C})-i\tilde{\Delta}}. 
\label{eq:av2}
\end{align}
From the equation, the conversion efficiency $\eta\u{SFG}$ is described by 
\begin{equation}
    \eta\u{SFG} = \frac{\braket{\hat{a}_{\mathrm{v},OUT}^\dagger\hat{a}_{\mathrm{v},OUT}}}{\braket{\hat{a}_{\mathrm{t},IN}^\dagger\hat{a}_{\mathrm{t},IN}}} = \frac{\tilde{\gamma\u{r}}\tilde{\alpha}P}{\frac{1}{4}(1+\tilde{\alpha}P)^2+\tilde{\Delta}^2}, 
    \label{eq:sfg}
\end{equation}
where $P$ is the pump power and 
$\tilde{\alpha}$ is a proportional constant satisfying 
$\tilde{C} = \tilde{\alpha}P$.
The conversion bandwidth around a fixed resonant frequency is 
$(1+\tilde{\alpha}P)\gamma\u{all}$. 

We extend the discussion to the case where frequency-multiplexed signal photons with the frequency spacing $\Delta \omega\u{comb}$ are injected to the converter.
We denote the $i$-th signal mode by $\hat{a}_{\U{t},i}$ and the $j$-th converted mode by $\hat{a}_{\U{v},j}$, with respective angular frequencies $\omega_{\mathrm{t},i}$ and $\omega_{\mathrm{v},j}$. 
The frequencies of the converted modes are determined by the resonant frequencies of the cavity, 
whereas the signal frequencies of SFG can be arbitrarily designed.
We consider the signal and converted frequencies such that when a signal mode $\hat{a}_{\U{t},i}$ is coupled to a converted mode $\hat{a}_{\U{v},j}$ by the pump light at the angular frequency $\omega_{\U{p},(i,j)}(=\omega_{\U{v},j}-\omega_{\U{t},i})$, while no other signal mode is effectively coupled to any converted mode, like depicted in Fig.~\ref{fig:theory_tweezer}~(b). 
Under the assumption, the theoretical treatment of the tweezing operation of the signal modes does not change from the previous paragraph.  
Then, the complex amplitudes of the converted and the remained modes after the tweezing process are described by the same forms of Eqs.~(\ref{eq:av}) and (\ref{eq:at}), with an appropriate reassignment of the modes and frequencies in the present context. 
When the pump frequency is shifted by the comb spacing $\Delta\omega\u{comb}$ of the input frequency comb from the setting shown in Fig.~\ref{fig:theory_tweezer}~(b), the selected input mode shifts by one frequency channel while being converted to the same frequency, which is shown in Fig.~\ref{fig:theory_tweezer}~(c). 
In contrast, when the pump frequency is shifted by the FSR $\Delta\omega\u{FSR}$ of the cavity, the selected input mode remains unchanged, while the converted frequency shifts to the neighboring cavity resonance, as shown in Fig.~\ref{fig:theory_tweezer}~(d). 
By summarizing these two characteristics, the energy conservation relation of the SFG can be written as 
\begin{align}
\omega_{\U{t},i+n} + (\omega_{\U{p},(i,j)}+m \Delta\omega\u{FSR}+n\Delta\omega\u{comb}) = \omega_{\U{v},j+m}, 
\label{eq:energyconservation}
\end{align}
where $\omega_{\U{t},i+n}=\omega_{\U{t},i} - n\Delta\omega\u{comb}$ and $\omega_{\U{v},j+m}=\omega_{\U{v},j}+m\Delta\omega\u{FSR}$. 
This equation demonstrates the dual capability of (i) selecting any input frequency bin as the target for conversion and (ii) routing a photon in the selected bin to any desired output mode, enabled by the appropriate choice of the pump frequency.
These capabilities offer full control over the input-output mapping in the frequency domain. 

In the above discussion with Fig.~\ref{fig:theory_tweezer}, 
we have implicitly assumed $\Delta\omega\u{comb}\ll \Delta\omega\u{FSR}$. 
However, even if the input $\omega_{\U{t},i+n}$ is replaced with 
$\omega_{\U{t},i+n}+k\Delta\omega\u{FSR}$, Eq.~(\ref{eq:energyconservation}) still holds by interpreting $m$ as $m+k$ for integer $k$. 
Therefore, the sequence of frequency-multiplexed input light is allowed to be $\{ \omega_{\U{t},i}+k_0\Delta\omega\u{FSR},\cdots, \omega_{\U{t},i+n}+k_n\Delta\omega\u{FSR}\}$ instead of $\{ \omega_{\U{t},i}, \cdots, \omega_{\U{t},i+n} \}$.

One important point to note is that the degree of frequency multiplexing of the signal light is limited by the performance of the cavity. 
We denote the finesse of the cold cavity without frequency conversion process by  $F\u{cold}=\Delta\omega\u{FSR}/\gamma\u{all}$. 
As shown in Eq.~(\ref{eq:sfg}), the cavity linewidth 
$(1 + \tilde{\alpha}P )\gamma\u{all}$
depends on the conversion efficiency. 
We consider the case of maximum conversion efficiency~($\tilde{\alpha}P=1$) as a representative scenario, in which the cavity linewidth becomes twice that of the cold cavity~$\gamma\u{all}$, resulting in the finesse being reduced by half to $F\u{cold}/2$. 
If the input frequency modes are arranged within a single FSR at a spacing of $\Delta\omega\u{comb}$, the number of frequency-multiplexed channels is limited by $N = \lfloor F\u{cold}/2 \rfloor$. 
An unexpected process, where a signal light in the frequency mode adjacent to the target mode is converted with non-negligible efficiency, decreases as the multiplexing degree is reduced.
Based on the above discussion, our implementation demonstrates the concept introduced in Sec.~\ref{sec:theory}\ref{subsec:design_concept}.

\section{Experiment}
\subsection{Experimental setup}
\begin{figure*}
    \centering
    \includegraphics[width = 0.8\linewidth]{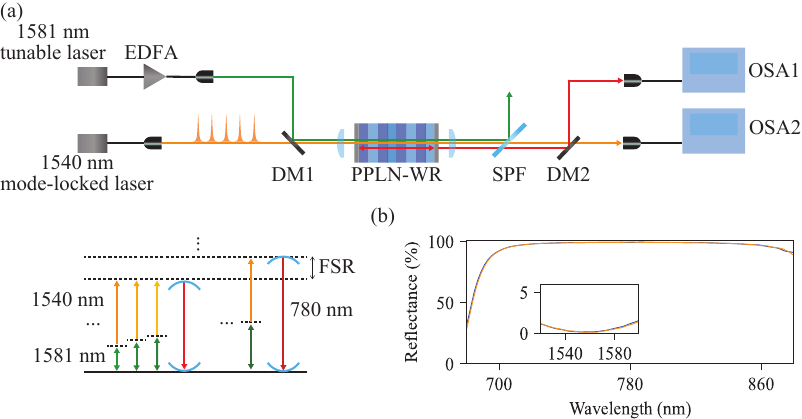}
    \caption{(a) Experimental setup for selective frequency up-conversion and the energy diagram of CS-QFC. EDFA: erbium-doped fiber amplifier, DM: dichroic mirror, SPF: short pass filter, OSA: optical spectral analyzer.
    (b)Reflectances of both ends of the PPLN-WR, shown in blue and orange.}
    \label{fig:experimental_setup}
\end{figure*}
Fig.~\ref{fig:experimental_setup} shows the experimental setup. 
The pump light around $\lambda_\mathrm{p}\sim$\SI{1581}{nm}
for the selective frequency up-conversion process is prepared 
using a frequency-tunable continuous wave~(cw) laser. 
It is amplified by an erbium-doped fiber amplifier~(EDFA), and then focused on the PPLN waveguide resonator~(PPLN-WR) using an aspheric lens with a focal length of \SI{8}{mm}. 
The frequency-multiplexed signal light is prepared by a mode-locked laser
with the center wavelength of \SI{1540}{nm}, 
the repetition rate of $\Delta\omega\u{comb}=2\pi\times $ \SI{1}{GHz}, 
and the linewidth of \SI{5}{ps} for each tooth. 
After being combined with the pump light at a dichroic mirror~(DM1), 
it is focused on the PPLN-WR. 

The length of the PPLN waveguide is $L^\mathrm{ex}=$\SI{19.67}{mm}. 
The end faces of the PPLN waveguide are flat polished, 
and coated by dielectric multilayers. 
The reflectance of each end face is \SI{99.5}{\%} around \SI{780}{nm} for high reflective coating. 
The corresponding value of $\gamma\u{r}$ in Eq.~(\ref{eq:sfg}) is 
$\gamma\u{r}^\U{ex}=\SI{6.0}{MHz}$. 
In contrast, those around \SI{1539}{nm} and \SI{1581}{nm} 
are less than \SI{0.4}{\%} for antireflective coatings. 
The wavelength dependency of the reflectance is shown in Fig.~\ref{fig:experimental_setup}~(b).

After the PPLN-WR, the pump light is eliminated using a short-pass filter~(SPF). 
The \SI{780}{nm} light generated at the PPLN-WR 
and the \SI{1539}{nm} light are separated by DM2. 
They are coupled to single-mode fibers followed by optical spectral analyzers~(OSAs). 
The resolutions of the OSAs used in the experiment are \SI{0.005}{\nano\meter} corresponding to the angular frequency of $2\pi\times$ \SI{600}{\mega\Hz} for \SI{1540}{nm} light and 
\SI{0.02}{\nano\meter} corresponding to $2\pi\times$\SI{10}{\giga\Hz} for \SI{780}{nm} light.

\subsection{Experimental results}
\begin{figure}
    \centering
    \includegraphics[width = \linewidth]{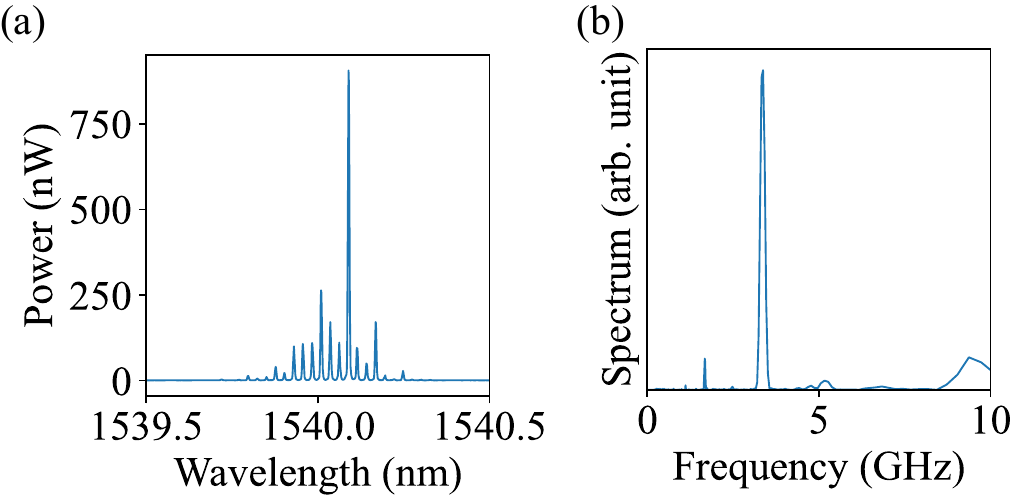}
    \caption{(a) The spectrum measured by the OSA2. 
    (b) The Fourier transform of the spectrum in (a). 
    The highest peak frequency is \SI{3.3}{\giga\Hz}.}
    \label{fig:fsr_estimation}
\end{figure}
\begin{figure}[t]
    \centering
    \includegraphics[width = \linewidth]{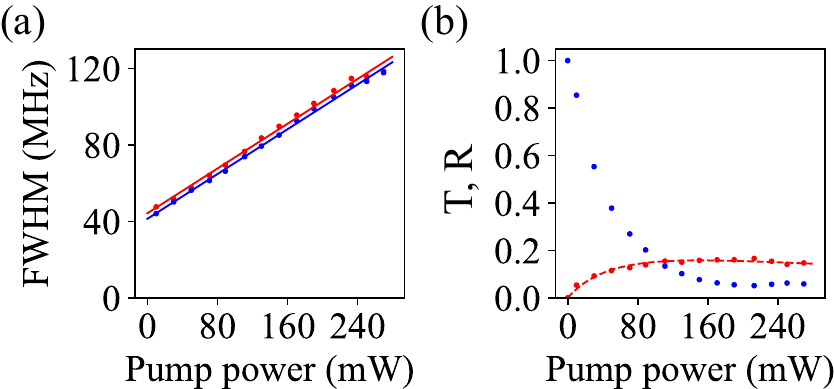}
    \caption{The pump power dependencies of (a) the bandwidth of the frequency conversion estimated from converted~(red) and signal~(blue) light and (b) conversion efficiency~($T$, red) and unconverted efficiency~($R$, blue). The dashed line is calculated by Eq.~(\ref{eq:sfg}) using 
    $\tilde{\alpha}^\U{ex} = \SI{7.3e-3}{\per\milli\watt}$ 
    and $\tilde{\gamma}\u{r}^\U{ex}=0.17$.}
    \label{fig:efficiency}
\end{figure}
Before performing the selective frequency up-conversion, 
we first performed a rough estimation of the FSR of the PPLN-WR for \SI{780}{nm}. 
For this, we input the frequency-multiplexed light 
at the center wavelength of \SI{1560}{nm} with the repetition rate of \SI{1}{GHz}, 
instead of the \SI{1540}{nm} signal light in Fig.~\ref{fig:experimental_setup}. 
The \SI{1560}{nm} light was frequency-doubled to the \SI{780}{nm} light 
by second harmonic generation~(SHG) in the PPLN-WR. 
The SHG processes of the frequency comb around \SI{1560}{nm} occur only when the second harmonic frequencies match near the cavity resonance. 
As a result, 
the comb spacing of the SHG light around \SI{780}{nm} should reflect the cavity FSR. 
Because the OSA for \SI{780}{nm} cannot resolve the comb spectrum due to the poor resolution, we converted the \SI{780}{nm} light to \SI{1540}{nm} 
by difference frequency generation~(DFG) with the pump light at \SI{1581}{nm} in the PPLN-WR. 
The observed spectrum of the DFG light at \SI{1540}{nm} is shown in Fig.~\ref{fig:fsr_estimation}~(a), 
in which we see the clear resonant structure. 
The Fourier transform of the spectrum is shown in Fig.~\ref{fig:fsr_estimation}~(b). 
From the highest peak, the cavity FSR for \SI{780}{nm} is estimated to be $\Delta\omega\u{FSR}=2\pi\times $\SI{3.3}{GHz}, which is consistent with previous reported values in PPLN-WRs for the telecom wavelengths with similar lengths~\cite{Tani2018,Yamazaki2022}.

Next, we characterized the conversion bandwidth and conversion efficiency around a fixed resonant frequency as the basic properties of the cavity-enhanced effect on the frequency up-conversion, similar to the reported methods in Refs.~\cite{ikuta2022optical,murakami2025low}. 
The experimental result of the pump power dependency of the conversion bandwidth is shown in Fig.~\ref{fig:efficiency}~(a). 
From the result, the conversion bandwidth is proportional to the pump power $P$ 
as predicted in Eq.~(\ref{eq:sfg}). 
By fitting the observed data using a function 
$(1 + \tilde{\alpha} P) \gamma\u{all}$, 
$\tilde{\alpha}^\U{ex} = \SI{7.3e-3}{\per\milli\watt}$ 
and $\gamma\u{all}^\U{ex} =2\pi\times \SI{40}{MHz}$ are estimated. 
In Fig.~\ref{fig:efficiency}~(b), 
we show the experimental result of the conversion efficiencies. 
In the figure, 
the maximum conversion efficiency was normalized to be 
$\tilde{\gamma}\u{r}^\U{ex}=\gamma\u{r}^\U{ex}/\gamma\u{all}^\U{ex}=0.17$ 
at the pump power of $P\u{max}=1/\tilde{\alpha}^\U{ex}=$\SI{140}{mW}.  
We see that the pump power dependency of the conversion efficiency is consistent with the theoretical curve in Eq.~(\ref{eq:sfg}) using the parameters estimated in Fig.~\ref{fig:efficiency}~(a). 

\begin{figure*}
    \centering
    \includegraphics[width = \linewidth]{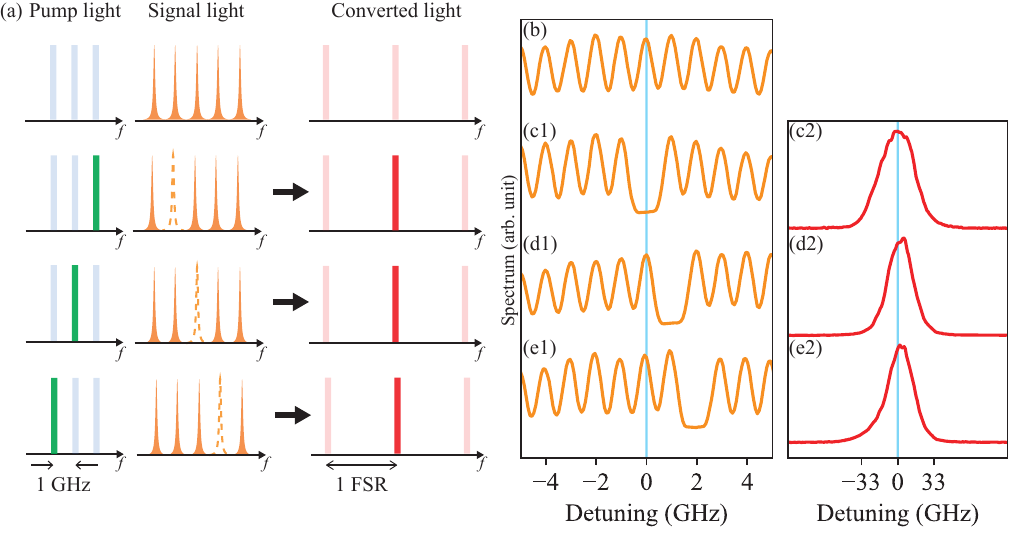}
    \caption{
    Selective frequency up-conversion, from light in different frequency modes to a fixed frequency mode. 
    (a) Concept. (b) Input signal light around \SI{1540}{nm}. 
    (c1), (d1), (e1) Observed signal light after frequency conversion using pump light with three different frequency detunings of $\omega\u{p,0}$ and $\omega\u{p,0}\pm \Delta\omega\u{comb}$. 
    (c2), (d2), (e2) Observed converted light around \SI{780}{nm} corresponding to (c1), (d1) and (e1), respectively.  
    }
    \label{fig:result_1GHz}
\end{figure*}
In the following, we performed the selective frequency up-conversion experiments.
We set the pump power to \SI{180}{\milli\watt} which gives the conversion bandwidth of \SI{92}{\mega\Hz}. 
We conducted two types of experiments: one demonstrating selective conversion of light in an arbitrary frequency bin among frequency-multiplexed channels, and the other demonstrating conversion of light in a specific frequency bin into an arbitrary resonant mode of the cavity. 
In the former experiment, we performed the frequency up-conversion by switching the pump frequencies among $\omega_{\mathrm{p},(0,0)}~(=2\pi \times \SI{189.542}{\tera\Hz})$ and 
$\omega_{\mathrm{p},(0,n)}=\omega_\mathrm{p,(0,0)} - n \Delta\omega\u{comb}$ $(n=1,2)$ around the wavelength of \SI{1581}{nm}, 
using only one at a time, as illustrated in Fig.~\ref{fig:result_1GHz}~(a). 
The observed spectra of the transmitted signal light around \SI{1540}{\nano\meter} 
and the converted light around \SI{780}{nm} are shown in Figs.~\ref{fig:result_1GHz}~(b) -- (e). 
When the pump frequency was $\omega_\mathrm{p,0}$, 
only the light in a single frequency bin centered at the frequency denoted by $\omega\u{t,0}$ was converted as shown in Fig.~\ref{fig:result_1GHz}~(c1). 
When the pump frequencies were set to $\omega\u{p,(0,0)} - \Delta\omega\u{comb}$ and $\omega\u{p,(0,0)} - 2\Delta\omega\u{comb}$, 
the frequency bin of the light to be converted shifts to a higher frequency bin by one and two bins, respectively, as in Figs.~\ref{fig:result_1GHz}~(d1) and (e1). 
In contrast, the spectra of the converted light around \SI{780}{nm} are found to have the same center frequency regardless of the pump frequencies, 
which are shown in Figs.~\ref{fig:result_1GHz}~(c2), (d2) and (e2). 
This is due to the energy conservation law of the SFG as 
$(\omega\u{p,(0,0)} \pm n\Delta\omega\u{comb}) + (\omega\u{t,0} \mp n\Delta\omega\u{comb}) = 
\omega\u{p,(0,0)}+\omega\u{t,0}$, which is independent of $n$.  
Considering that frequency bins other than the target are not converted at all, 
the frequency up-conversion with the ability to select the input mode was shown. 

\begin{figure*}
    \centering
        \includegraphics[width = \linewidth]{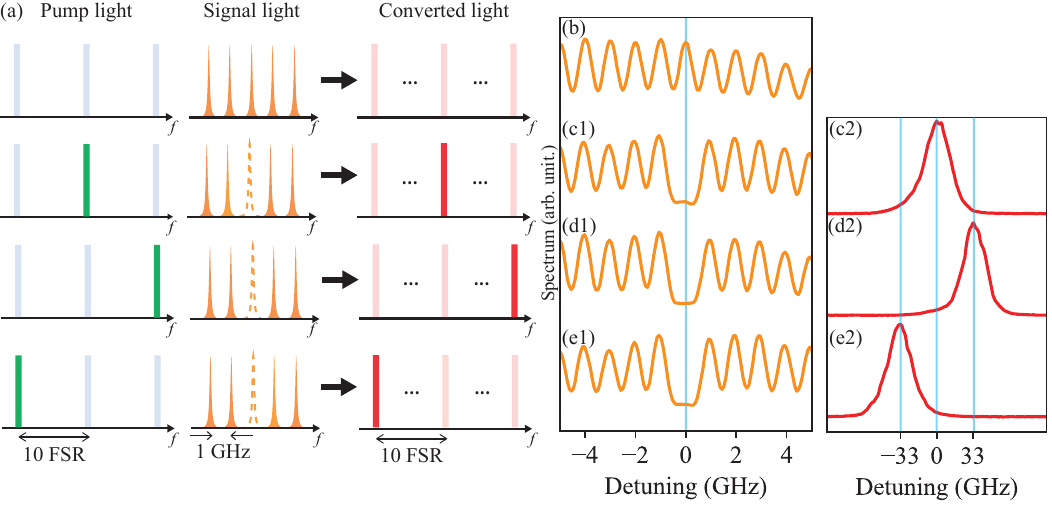}
    \caption{
    Selective frequency up-conversion, from light in a fixed frequency mode to different frequency modes. 
    (a) Concept. (b) Input signal light around \SI{1540}{nm}. 
    (c1), (d1), (e1) Observed signal light after frequency conversion using pump light with three different frequency detunings of $\omega\u{p,0}$ and $\omega\u{p,0}\pm 10\Delta\omega\u{FSR}$. 
    (c2), (d2), (e2) Observed converted light around \SI{780}{nm} corresponding to (c1), (d1) and (e1), respectively.  }
    \label{fig:result_10FSR}
\end{figure*}
Next, to demonstrate the selectability of the converted mode, we performed the frequency up-conversion by switching the pump frequencies of 
$\omega\u{p,(0,0)}$ and $\omega\u{p,(0,\pm 10)}=\omega\u{p,(0,0)} \pm 10\Delta\omega\u{FSR}$, 
as illustrated in Fig.~\ref{fig:result_10FSR}~(a). 
The reason for the far detuning of their frequencies is the limitation of the resolution \SI{10}{\giga\Hz} of OSA for the \SI{780}{nm} light. 
The observed spectra of the transmitted and converted light are shown in Figs.~\ref{fig:result_10FSR}~(b) -- (e). 
From the spectra of the transmitted light in Figs.~\ref{fig:result_10FSR}~(c1), (d1), and (e1), we see that, regardless of the pump frequencies, the frequency bin to be converted around \SI{1540}{nm} did not change. 
In contrast, in Figs.~\ref{fig:result_10FSR}~(c2), (d2) and (e2), 
the central frequencies of the converted light around \SI{780}{nm} were shifted by $\pm 10\Delta\omega\u{FSR}=\pm 2\pi\times\SI{33}{\giga\Hz}$ when the pump frequencies of $\omega\u{p,(0,0)} \pm 10\Delta\omega\u{FSR}$ were used. 
The results show that the frequency up-conversion enables the selection of one of the cavity resonant modes as the converted mode. 
Together with the result described in the previous paragraph, we successfully demonstrated the selective frequency up-conversion 
from any frequency mode of the frequency-multiplexed light to any resonant mode of the cavity. 

\section{Discussion}\label{sec:discussion}
\subsection{Effect of the frequency tweezers on signal-to-noise ratio for frequency-multiplexed photons}
For practical quantum information applications, it is crucial that the optical frequency tweezers exhibit a sufficiently large SNR for a single-photon input. 
While the performance of QFC in a single frequency mode has been well studied, its performance for frequency-multiplexed photons is still not well investigated. 
In the following, we discuss this aspect in detail through an analysis of the effect of the frequency tweezers on the SNR for the frequency-multiplexed input photons. 

We first describe the amount of the noise photons. 
In QFC processes between \SI{780}{nm} and the telecom wavelengths, as in our experiment, it is known that the dominant noise source is anti-Stokes~(AS) photons generated by the strong pump light used for QFC. 
In Ref.~\cite{murakami2023quantum}, 
the amount of the AS photons around \SI{1540}{nm} produced by the \SI{1580}{nm} pump light was reported using normalized coefficient $n_{1540}^\U{ref}=\SI{25}{cps/(\milli\watt~\giga\Hz~\milli\meter)}$ just after the PPLN-WR. 
The noise photons around \SI{780}{nm} are generated as a result of SFG of the AS photons, described using Eq.~(\ref{eq:av}). 
In the time domain, the AS photons are continuously broadened due to the use of the cw pump light.  
As a result, the total number of noise photons produced around a single resonant frequency at \SI{780}{nm} is
\begin{align}
    N\u{780} &= n_{1540}^\U{ref}L^\U{ex}PT\u{w}\int_{-\infty}^{\infty}\frac{\tilde{\gamma\u{r}}\tilde{\alpha}P}{(\frac{1+\tilde{\alpha}P}{2})^2 +\tilde{\Delta}^2}d\Delta \nonumber\\
    &= n_{1540}^\U{ref}L^\U{ex}T\u{w}\frac{2\pi\gamma\u{r}\tilde{\alpha}P^2}{1+\tilde{\alpha}P}, 
\end{align}
where $T\u{w}$ is the detection time window. 
At the maximum conversion efficiency with $P=P\u{max}(=1/\tilde{\alpha})$, we obtain
\begin{equation}
    N\u{780} = \pi \tilde{\gamma\u{r}}n_{1540}^\U{ref}L^\U{ex}T\u{w}\gamma\u{all}P\u{max}.
\end{equation}

Next, for estimating the amount of the signal photon after the tweezing, 
we assume each tooth of the frequency-multiplexed signal photons has the time-reversed Lorentzian decay shape to enable efficient coupling to the cavity mode~\cite{Aljunid2013}, and its linewidth is $n$ times narrower than that of the cavity.
Due to the finite detection time window of $T\u{w}$ and the cavity structure, the frequency-converted photon is filtered in both temporal and spectral domains. 
From Eq.~(\ref{eq:av}) with $\tilde{C}=1$, the detected photon number of the converted photon around the single resonant frequency with the time window $T\u{w}$  becomes 
\begin{align}
    S\u{780} =& \tilde{\gamma}\u{r}N\u{in}\int_{T\u{w}} \left|\mathcal{F}\left[\frac{e^{i\phi}\sqrt{\tilde{\gamma\u{r}}}}{1 - i\tilde{\Delta}}\cdot \frac{1}{\sqrt{\pi \gamma\u{all}}}\frac{\sqrt{1/n}}{1/n + i\tilde{\Delta}}\right]\right|^2\mathrm{d}t \\
    =& \tilde{\gamma}\u{r}N\u{in}\frac{n}{1+n}\left(1-e^{-\frac{2T\u{w}\gamma\u{all}}{1+n}}\right),
\end{align}
where $N\u{in}$ is the photon number in the single tooth of the input signal and $\mathcal{F}[\cdot]$ represents the Fourier transform. 

From $N_{780}$ and $S_{780}$ calculated above, the SNR denoted by $\zeta$ is described as  
\begin{equation}
    \zeta = \frac{S_{780}}{N_{780}}. 
\end{equation}
This leads to the relationship between the autocorrelation functions $g^{(2)}\u{in}$ and $g^{(2)}\u{out}$ of the input photon and converted photon in one tooth as~\cite{ikuta2014frequency}
\begin{equation}
    g^{(2)}\u{out} = \frac{1}{(1+\zeta)^2}\left(g\u{in}^{(2)}\zeta^2 + 1 + 2\zeta\right),
    \label{eq:g2}
\end{equation}
where the autocorrelation function of the AS photons is assumed to be 1. 
As an example, for $N\u{in}=0.1$, $g\u{in}^{(2)} = 0.01$, $n=5$ and $T\u{w}=\SI{125}{\nano\second}$, 
with experimental values of $L^\U{ex}=$\SI{19.67}{mm}, $P\u{max}^\U{ex}=$\SI{140}{mW}, and $\gamma\u{all}^\U{ex}=2\pi\times$\SI{40}{\mega\Hz}, we obtain $\zeta= 62.5$, resulting in $g\u{out}^{(2)}(0) = 0.05$.
This value indicates that a single tweezing process can be performed with negligible degradation of the autocorrelation function. 

\begin{figure}
    \centering
    \includegraphics[width = \linewidth]{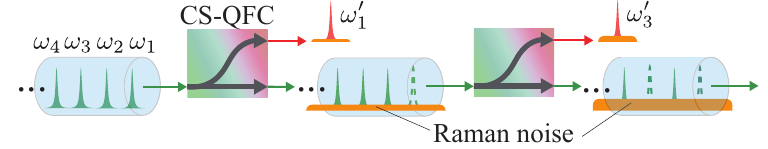}
    \caption{The concept of individual access for each tooth by selective frequency up-conversion. 
    The recursive access generates the AS photons and accumulates their number in unextracted teeth.}
    \label{fig:indivisual_SFG}
\end{figure}
\begin{figure}[t]
    \centering
    \includegraphics[width = \linewidth]{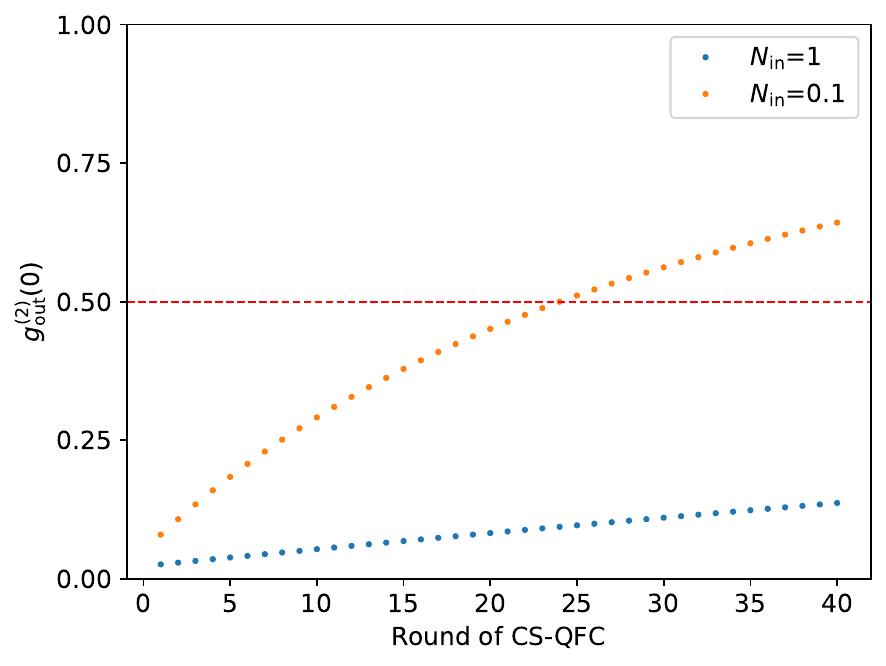}
    \caption{The relationship between the extracted round of the photons by CS-QFC and auto-correlation function $g^{(2)}\u{out}$ for the initial $g^{(2)}$ value of 0.01. $N\u{in}$ describes the mean photon number in each tooth.}
    \label{fig:snr_estimation}
\end{figure}

When we sequentially convert each component of an $M$-fold frequency-multiplexed light into distinct output modes, the $g^{(2)}$ value degrades with each QFC step, like shown in Fig.~\ref{fig:indivisual_SFG}, 
because noise photons accumulate in the frequency modes that are not converted. As a result, the $g^{(2)}$ value of the converted photons through the final round of QFC is obtained by applying Eq.~(\ref{eq:g2}) with substituting $\zeta/M$ to $\zeta$.
In this model, Fig.~\ref{fig:snr_estimation} shows the relationship between the number of the multiplexed frequency channels and $g^{(2)}\u{out}$ obtained after the final round of QFC. 
As mentioned in Sec.~\ref{sec:theory}, the present experimental parameters allow for multiplexing up to $F\u{cold}/2 \sim 40$ channels, 
meaning that the maximum number of rounds for the CS-QFC is $40$. 
From Fig.~\ref{fig:snr_estimation}, for $N\u{in}=0.1$, $g^{(2)}\u{out}$ is well below 1, which is the lower bound of the classical wave theory, after the final round of CS-QFC. Furthermore, after 23 rounds, $g^{(2)}\u{out}$ remains below $0.5$, indicating that the extracted photon stays within the single-photon regime. 
For $N\u{in}=1$, even after the final round of the tweezing operations, the value of $g^{(2)}\u{out}$ still remains below 0.5. 
These results imply that our method will be applicable to the extraction and delivery of entanglement using frequency-multiplexed photons.

\subsection{Use cases of the channel-selective quantum frequency converter}
\label{subsec:4b}
We discuss four use cases of the CS-QFC in quantum information applications. 
The first application is the frequency-selective BSM described in Sec.~\ref{intro} with Fig.~\ref{fig:concept}. 
In a reconfigurable quantum network, 
two distributed photons entangled with quantum systems at two end nodes are not necessarily in the same frequency channel, even though the end nodes are intended to be entangled. In such cases, CS-QFCs based on the optical frequency tweezers enable the conversion of the two different-color photons to the same frequency like Ref.~\cite{Takesue2008}, without disturbing photons in other modes, thereby allowing a successful BSM to be performed. 

\begin{figure}[t]
    \centering
    \includegraphics[width = \linewidth]{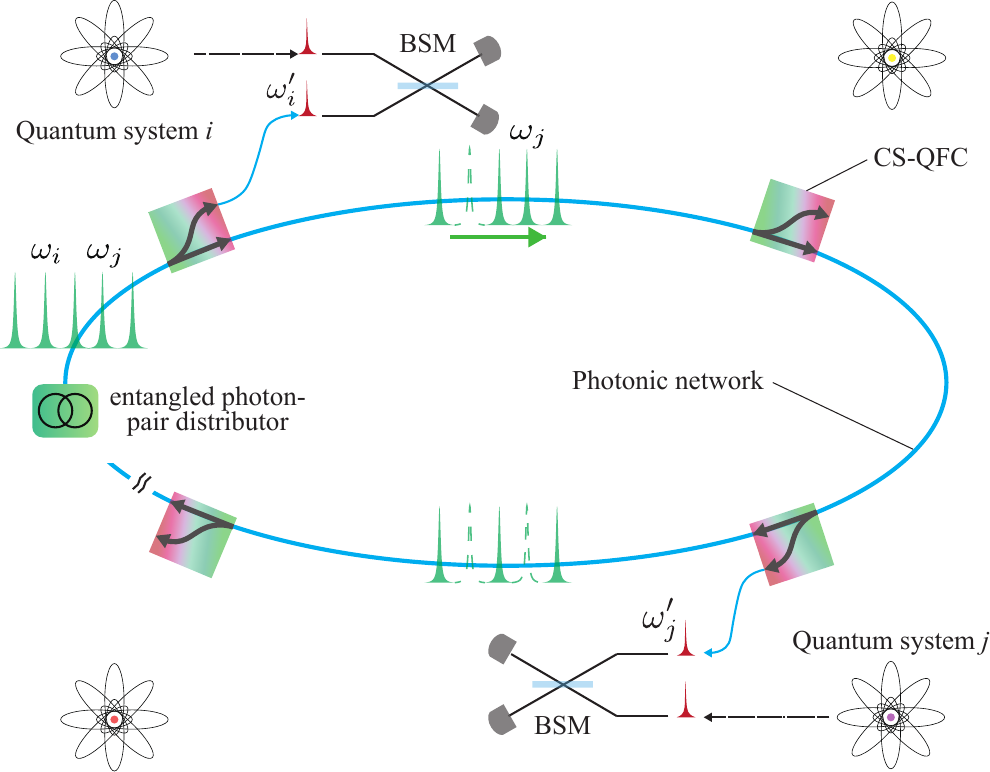}
    \caption{The concept of photonic quantum network with CS-QFCs.}
    \label{fig:concept_tweezer_network}
\end{figure}
The second application involves a ring-shaped photonic quantum network, in which we assume the existence of an entanglement distributor that supplies frequency-multiplexed entangled photon pairs at the telecom band like Refs.~\cite{wengerowsky2018entanglement,joshi2020trusted,qi202115,fujimoto2022entanglement}, as illustrated in Fig.~\ref{fig:concept_tweezer_network}. 
Two end nodes, denoted by $i$ and $j$, to be entangled perform CS-QFCs at their nodes and then share the entangled photon pair. 
Entanglement between quantum systems located at the two nodes can be established as needed by performing a BSM at each node between a photon emitted from the local system and the photon selectively picked up by the CS-QFC.
If another pair of nodes other than $i$ and $j$ wishes to do the same, they can extract a remaining entangled photon pair from the frequency-multiplexed resource using the CS-QFC. 

The third application lies in the frequency multiplexed cavity-QED systems.
Recent studies have reported both theoretical proposals~\cite{kikura2025passive,ji2025global} and experimental demonstration~\cite{hu2025site} of selective coupling between neutral atoms and cavity resonances in cavity-QED systems.
In these works, selective coupling is achieved by inducing an AC Stark shift with a shifting laser, shifting the energy levels of non-target atoms away from the cavity resonance.
Additionally, Ref.~\cite{kikura2025passive} proposes that tuning the shifts of atoms to integer multiples of the cavity FSR enables the realization of a frequency-multiplexed quantum memory within a single cavity, where different atoms are assigned distinct shifts to couple with different cavity resonances.
Ref.~\cite{hu2025site} reports energy-level shifts on the order of GHz, indicating that such shifts are experimentally feasible for a frequency-multiplexed quantum memory with GHz-scale FSR.
Our CS-QFC not only enables frequency-selective conversion of input photons but also provides precise control over the output frequency.
By matching the FSR of the cavity used for cavity-QED with that of the CS-QFC cavity, both selective extraction of individual photons from multiplexed modes and controlled access to specific atoms can be simultaneously achieved by CS-QFC.

The last case is related to the second example. 
Up to this point, our discussion has focused on frequency up-conversion from telecom to visible wavelengths. 
However, from Eq.~(\ref{eq:at}), for the input of visible light, we obtain 
\begin{align}
\hat{a}_{\U{t},OUT}
&= \frac{e^{-i\phi}\sqrt{\tilde{\gamma\u{r}}\tilde{C}}}{\frac{1}{2}(1+\tilde{C})-i\tilde{\Delta}}\hat{a}_{\U{v},IN}, 
\end{align}
when there is no input from the corresponding telecom light as $\hat{a}_{\U{t},IN}=0$. 
This equation indicates that, as in the case of frequency up-conversion, the reverse process is also possible.
Therefore, in Fig.~\ref{fig:concept_tweezer_network}, if there is an empty channel without telecom photons within the frequency-multiplexed channels, it is possible to convert visible photons injected from the end node of the network into that channel without affecting telecom photons in other frequency channels.
While the second example demonstrates the function as a drop filter, in contrast, this example employs the function as an add filter through the frequency conversion. 

\section{Conclusion}
We have demonstrated channel-selective frequency up-conversion of frequency-multiplexed light from \SI{1540}{\nano\meter} to \SI{780}{\nano\meter}. 
Using a PPLN-WR with a resonant structure around \SI{780}{nm}, and designing the mode spacings of the cavity and the input light, the device selectively convert the light in a specific frequency mode without affecting other frequency modes. 
The device is also capable of routing the selected signal to a desired converted mode defined by the cavity resonance structures. 
In the experiment, we successfully demonstrated both the channel selectivity and the routing capability. 
Based on the experimental results, we have numerically confirmed the feasibility of this method from the perspective of the amount of noise photons.
In addition, we discussed the applicability of the device as a CS-QFC by introducing use cases, including BSMs between photons originating from different frequency channels 
in the context of frequency-multiplexed quantum repeaters. 
Additional examples showed that the device can operate as a QFC-based add/drop filter, 
similar to those used in classical optical fiber networks employing ROADMs. 
This technology demonstrated here is expected to serve a valuable building block for designing flexible frequency-multiplexed quantum networks that integrate multiple quantum matter systems interacting with the photons. 

\section*{Funding}
This work was supported by Moonshot R \& D, JST JPMJMS2066, JST JPMJMS226C; FOREST Program, JST JPMJFR222V; R \& D of ICT Priority Technology Project JPMI00316, Asahi Glass Foundation, JSPS JP22J20801, and Program for Leading Graduate Schools: Interactive Materials Science Cadet Program.

\section*{Acknowledgments}
T.Y. and R.I. acknowledge the members of the Quantum Internet Task Force for the comprehensive and interdisciplinary discussions on the quantum internet.
The authors thank Seigo Kikura for his insightful suggestions regarding the application of CS-QFC.

\bibliography{sample}

@article{van2022entangling,
  title={Entangling single atoms over 33 km telecom fibre},
  author={van Leent, Tim and Bock, Matthias and Fertig, Florian and Garthoff, Robert and Eppelt, Sebastian and Zhou, Yiru and Malik, Pooja and Seubert, Matthias and Bauer, Tobias and Rosenfeld, Wenjamin and others},
  journal={Nature},
  volume={607},
  number={7917},
  pages={69--73},
  year={2022},
  publisher={Nature Publishing Group UK London}
}

@article{tchebotareva2019entanglement,
  title={Entanglement between a diamond spin qubit and a photonic time-bin qubit at telecom wavelength},
  author={Tchebotareva, Anna and Hermans, Sophie LN and Humphreys, Peter C and Voigt, Dirk and Harmsma, Peter J and Cheng, Lun K and Verlaan, Ad L and Dijkhuizen, Niels and De Jong, Wim and Dr{\'e}au, Ana{\"\i}s and others},
  journal={Physical Review Letters},
  volume={123},
  number={6},
  pages={063601},
  year={2019},
  publisher={APS}
}

@article{bersin2024telecom,
  title={Telecom networking with a diamond quantum memory},
  author={Bersin, Eric and Sutula, Madison and Huan, Yan Qi and Suleymanzade, Aziza and Assumpcao, Daniel R and Wei, Yan-Cheng and Stas, Pieter-Jan and Knaut, Can M and Knall, Erik N and Langrock, Carsten and others},
  journal={PRX Quantum},
  volume={5},
  number={1},
  pages={010303},
  year={2024},
  publisher={APS}
}

@article{bock2018high,
  title={High-fidelity entanglement between a trapped ion and a telecom photon via quantum frequency conversion},
  author={Bock, Matthias and Eich, Pascal and Kucera, Stephan and Kreis, Matthias and Lenhard, Andreas and Becher, Christoph and Eschner, J{\"u}rgen},
  journal={Nature Communications},
  volume={9},
  number={1},
  pages={1998},
  year={2018},
  publisher={Nature Publishing Group UK London}
}

@article{wehner2018quantum,
  title={Quantum internet: A vision for the road ahead},
  author={Wehner, Stephanie and Elkouss, David and Hanson, Ronald},
  journal={Science},
  volume={362},
  number={6412},
  pages={eaam9288},
  year={2018},
  publisher={American Association for the Advancement of Science}
}

@article{kimble2008quantum,
  title={The quantum internet},
  author={Kimble, H Jeff},
  journal={Nature},
  volume={453},
  number={7198},
  pages={1023--1030},
  year={2008},
  publisher={Nature Publishing Group}
}

@article{kumar1990quantum,
  title={Quantum frequency conversion},
  author={Kumar, Prem},
  journal={Optics letters},
  volume={15},
  number={24},
  pages={1476--1478},
  year={1990},
  publisher={Optica Publishing Group}
}

@article{ikuta2014frequency,
  title={Frequency down-conversion of 637 nm light to the telecommunication band for non-classical light emitted from NV centers in diamond},
  author={Ikuta, Rikizo and Kobayashi, Toshiki and Yasui, Shuto and Miki, Shigehito and Yamashita, Taro and Terai, Hirotaka and Fujiwara, Mikio and Yamamoto, Takashi and Koashi, Masato and Sasaki, Masahide and others},
  journal={Optics express},
  volume={22},
  number={9},
  pages={11205--11214},
  year={2014},
  publisher={Optica Publishing Group}
}

@article{ikuta2018polarization,
  title={Polarization insensitive frequency conversion for an atom-photon entanglement distribution via a telecom network},
  author={Ikuta, Rikizo and Kobayashi, Toshiki and Kawakami, Tetsuo and Miki, Shigehito and Yabuno, Masahiro and Yamashita, Taro and Terai, Hirotaka and Koashi, Masato and Mukai, Tetsuya and Yamamoto, Takashi and others},
  journal={Nature Communications},
  volume={9},
  number={1},
  pages={1997},
  year={2018},
  publisher={Nature Publishing Group UK London}
}

@article{murakami2023quantum,
  title={Quantum frequency conversion using 4-port fiber-pigtailed PPLN module},
  author={Murakami, Shoichi and Fujimoto, Rintaro and Kobayashi, Toshiki and Ikuta, Rikizo and Inoue, Asuka and Umeki, Takeshi and Miki, Shigehito and China, Fumihiro and Terai, Hirotaka and Kasahara, Ryoichi and others},
  journal={Optics Express},
  volume={31},
  number={18},
  pages={29271--29279},
  year={2023},
  publisher={Optica Publishing Group}
}

@article{ikuta2022optical,
  title={Optical frequency tweezers},
  author={Ikuta, Rikizo and Yokota, Masayo and Kobayashi, Toshiki and Imoto, Nobuyuki and Yamamoto, Takashi},
  journal={Physical Review Applied},
  volume={17},
  number={3},
  pages={034012},
  year={2022},
  publisher={APS}
}

@article{ikuta2011wide,
  title={Wide-band quantum interface for visible-to-telecommunication wavelength conversion},
  author={Ikuta, Rikizo and Kusaka, Yoshiaki and Kitano, Tsuyoshi and Kato, Hiroshi and Yamamoto, Takashi and Koashi, Masato and Imoto, Nobuyuki},
  journal={Nature Communications},
  volume={2},
  number={1},
  pages={537},
  year={2011},
  publisher={Nature Publishing Group UK London}
}

@article{Aljunid2013,
  title = {Excitation of a Single Atom with Exponentially Rising Light Pulses},
  author = {Aljunid, Syed Abdullah and Maslennikov, Gleb and Wang, Yimin and Dao, Hoang Lan and Scarani, Valerio and Kurtsiefer, Christian},
  journal = {Phys. Rev. Lett.},
  volume = {111},
  issue = {10},
  pages = {103001},
  numpages = {5},
  year = {2013},
  month = {Sep},
  publisher = {American Physical Society},
  doi = {10.1103/PhysRevLett.111.103001},
  url = {https://link.aps.org/doi/10.1103/PhysRevLett.111.103001}
}

@article{Liu2024,
  title={Reconfigurable entanglement distribution network based on pump management of a spontaneous four-wave mixing source},
  author={Liu, Jingyuan and Liu, Dongning and Jin, Zhanping and Lin, Zhihao and Li, Hao and You, Lixing and Feng, Xue and Liu, Fang and Cui, Kaiyu and Zhang, Wei and others},
  journal={Science Advances},
  volume={10},
  number={50},
  pages={eado9822},
  year={2024},
  publisher={American Association for the Advancement of Science}
}

@article{Shapourian2025,
  title={Quantum Data Center Infrastructures: A Scalable Architectural Design Perspective},
  author={Shapourian, Hassan and Kaur, Eneet and Sewell, Troy and Zhao, Jiapeng and Kilzer, Michael and Kompella, Ramana and Nejabati, Reza},
  journal={arXiv preprint arXiv:2501.05598},
  year={2025}
}

@article{Sakuma2024,
  title={An optical interconnect for modular quantum computers},
  author={Sakuma, Daisuke and Taherkhani, Amin and Tsuno, Tomoki and Sasaki, Toshihiko and Shimizu, Hikaru and Teramoto, Kentaro and Todd, Andrew and Ueno, Yosuke and Hajdu{\'L}{\k{A}}ek, Michal and Ikuta, Rikizo and others},
  journal={arXiv preprint arXiv:2412.09299},
  year={2024}
}

@article{Chang2025,
title = {Recent advances in high-dimensional quantum frequency combs},
journal = {Newton},
volume = {1},
number = {1},
pages = {100024},
year = {2025},
issn = {2950-6360},
doi = {https://doi.org/10.1016/j.newton.2025.100024},
url = {https://www.sciencedirect.com/science/article/pii/S2950636025000167},
author = {Kai-Chi Chang and Xiang Cheng and Murat Can Sarihan and Chee Wei Wong},
}

@article{Azuma2023,
  title={Quantum repeaters: From quantum networks to the quantum internet},
  author={Azuma, Koji and Economou, Sophia E and Elkouss, David and Hilaire, Paul and Jiang, Liang and Lo, Hoi-Kwong and Tzitrin, Ilan},
  journal={Reviews of Modern Physics},
  volume={95},
  number={4},
  pages={045006},
  year={2023},
  publisher={APS}
}

@article{Yan2025,
  title={Ten-channel Hong--Ou--Mandel interference between independent optical combs},
  author={Yan, Wenhan and Hu, Yang and Du, Yifeng and Wang, Kai and Lu, Yan-Qing and Zhu, Shining and Ma, Xiao-Song},
  journal={Chinese Optics Letters},
  volume={23},
  number={4},
  pages={042701},
  year={2025}
}

@article{Huang2025,
  title={Massively parallel Hong-Ou-Mandel interference based on independent soliton microcombs},
  author={Huang, Long and Wang, Weiqiang and Wang, Fangxiang and Wang, Yang and Zou, Changling and Tang, LinHan and Little, Brent E and Zhao, Wei and Han, Zhengfu and Yang, Jun and others},
  journal={Science Advances},
  volume={11},
  number={5},
  pages={eadq8982},
  year={2025},
  publisher={American Association for the Advancement of Science}
}

@article{Ichihara2023,
  title={Frequency-multiplexed hong-ou-mandel interference},
  author={Ichihara, Mayuka and Yoshida, Daisuke and Hong, Feng-Lei and Horikiri, Tomoyuki},
  journal={Physical Review A},
  volume={107},
  number={3},
  pages={032608},
  year={2023},
  publisher={APS}
}

@article{Takesue2008,
  title = {Erasing Distinguishability Using Quantum Frequency Up-Conversion},
  author = {Takesue, Hiroki},
  journal = {Phys. Rev. Lett.},
  volume = {101},
  issue = {17},
  pages = {173901},
  numpages = {4},
  year = {2008},
  month = {Oct},
  publisher = {American Physical Society},
  doi = {10.1103/PhysRevLett.101.173901},
  url = {https://link.aps.org/doi/10.1103/PhysRevLett.101.173901}
}

@article{tanzilli2005photonic,
  title={A photonic quantum information interface},
  author={Tanzilli, Sebastien and Tittel, Wolfgang and Halder, Matthaeus and Alibart, Olivier and Baldi, Pascal and Gisin, Nicolas and Zbinden, Hugo},
  journal={Nature},
  volume={437},
  number={7055},
  pages={116--120},
  year={2005},
  publisher={Nature Publishing Group UK London}
}

@article{wengerowsky2018entanglement,
  title={An entanglement-based wavelength-multiplexed quantum communication network},
  author={Wengerowsky, S{\"o}ren and Joshi, Siddarth Koduru and Steinlechner, Fabian and H{\"u}bel, Hannes and Ursin, Rupert},
  journal={Nature},
  volume={564},
  number={7735},
  pages={225--228},
  year={2018},
  publisher={Nature Publishing Group UK London}
}

@article{Tani2018,
title = {Frequency comb generation in a quadratic nonlinear waveguide resonator},
author = {Rikizo Ikuta and Motoki Asano and Ryoya Tani and Takashi Yamamoto and Nobuyuki Imoto},
url = {https://www.osapublishing.org/oe/abstract.cfm?uri=oe-26-12-15551},
doi = {10.1364/OE.26.015551},
year = {2018},
date = {2018-06-06},
journal = {Optics Express},
volume = {26},
number = {12},
pages = {15551--15558},
pubstate = {published},
tppubtype = {article}
}

@article{Yamazaki2022,
title = {Massive-mode polarization entangled biphoton frequency comb},
author = {Tomohiro Yamazaki and Rikizo Ikuta and Toshiki Kobayashi and Shigehito Miki and Fumihiro China and Hirotaka Terai and Nobuyuki Imoto and Takashi Yamamoto},
url = {https://www.nature.com/articles/s41598-022-12691-7},
doi = {https://doi.org/10.1038/s41598-022-12691-7},
year = {2022},
date = {2022-05-28},
urldate = {2022-05-28},
journal = {Scientific Reports},
volume = {12},
number = {8964},
pages = {1 -- 8},
pubstate = {published},
tppubtype = {article}
}

@article{fujimoto2022entanglement,
  title={Entanglement distribution using a biphoton frequency comb compatible with DWDM technology},
  author={Fujimoto, Rintaro and Yamazaki, Tomohiro and Kobayashi, Toshiki and Miki, Shigehito and China, Fumihiro and Terai, Hirotaka and Ikuta, Rikizo and Yamamoto, Takashi},
  journal={Optics Express},
  volume={30},
  number={20},
  pages={36711--36716},
  year={2022},
  publisher={Optica Publishing Group}
}

@article{hattori1999plc,
  title={PLC-based optical add/drop switch with automatic level control},
  author={Hattori, Kuninori and Fukui, Masaki and Jinno, Masahiko and Oguma, Manabu and Oguchi, Kimio},
  journal={Journal of lightwave technology},
  volume={17},
  number={12},
  pages={2562},
  year={1999},
  publisher={IEEE}
}

@article{ford1999wavelength,
  title={Wavelength add-drop switching using tilting micromirrors},
  author={Ford, Joseph E and Aksyuk, Vladimir A and Bishop, David J and Walker, James A},
  journal={Journal of lightwave technology},
  volume={17},
  number={5},
  pages={904},
  year={1999},
  publisher={IEEE}
}

@article{ertel2006design,
  title={Design and performance of a reconfigurable liquid-crystal-based optical add/drop multiplexer},
  author={Ertel, J and Helbing, R and Hoke, C and Landolt, O and Nishimura, K and Robrish, P and Trutna, R},
  journal={Journal of lightwave technology},
  volume={24},
  number={4},
  pages={1674--1680},
  year={2006},
  publisher={IEEE}
}

@article{wang2021proposal,
  title={Proposal and proof-of-principle demonstration of fast-switching broadband frequency shifting for a frequency-multiplexed quantum repeater},
  author={Wang, Peng-Cheng and Pietx-Casas, Oriol and Falamarzi Askarani, Mohsen and Do Amaral, Gustavo Castro},
  journal={Journal of the Optical Society of America B},
  volume={38},
  number={4},
  pages={1140--1146},
  year={2021},
  publisher={Optical Society of America}
}

@article{arizono20241xn,
  title={{1xN DWDM channel selective quantum frequency conversion}},
  author={Arizono, Tomoaki and Kobayashi, Toshiki and Miki, Shigehito and Terai, Hirotaka and Kodama, Tsuyoshi and Shimoi, Hideki and Yamamoto, Takashi and Ikuta, Rikizo},
  journal={arXiv preprint arXiv:2409.08025},
  year={2024}
}

@article{murakami2025low,
  title={Low-noise quantum frequency conversion with cavity enhancement of the converted mode},
  author={Murakami, Shoichi and Kobayashi, Toshiki and Miki, Shigehito and Terai, Hirotaka and Kodama, Tsuyoshi and Sawaya, Tsuneaki and Ohtomo, Akihiko and Shimoi, Hideki and Yamamoto, Takashi and Ikuta, Rikizo},
  journal={Optica Quantum},
  volume={3},
  number={1},
  pages={55--63},
  year={2025},
  publisher={Optica Publishing Group}
}

@ARTICLE{Yu2020-va,
  title     = "Entanglement of two quantum memories via fibres over dozens of
               kilometres",
  author    = "Yu, Yong and Ma, Fei and Luo, Xi-Yu and Jing, Bo and Sun,
               Peng-Fei and Fang, Ren-Zhou and Yang, Chao-Wei and Liu, Hui and
               Zheng, Ming-Yang and Xie, Xiu-Ping and Zhang, Wei-Jun and You,
               Li-Xing and Wang, Zhen and Chen, Teng-Yun and Zhang, Qiang and
               Bao, Xiao-Hui and Pan, Jian-Wei",
  journal   = "Nature",
  publisher = "Springer Science and Business Media LLC",
  volume    =  578,
  number    =  7794,
  pages     = "240--245",
  month     =  feb,
  year      =  2020,
  language  = "en"
}

@ARTICLE{Liu2024-oz,
  title     = "Creation of memory-memory entanglement in a metropolitan quantum
               network",
  author    = "Liu, Jian-Long and Luo, Xi-Yu and Yu, Yong and Wang, Chao-Yang
               and Wang, Bin and Hu, Yi and Li, Jun and Zheng, Ming-Yang and
               Yao, Bo and Yan, Zi and Teng, Da and Jiang, Jin-Wei and Liu,
               Xiao-Bing and Xie, Xiu-Ping and Zhang, Jun and Mao, Qing-He and
               Jiang, Xiao and Zhang, Qiang and Bao, Xiao-Hui and Pan, Jian-Wei",
  journal   = "Nature",
  publisher = "Springer Science and Business Media LLC",
  volume    =  629,
  number    =  8012,
  pages     = "579--585",
  month     =  may,
  year      =  2024,
  language  = "en"
}

@article{joshi2020trusted,
  title={A trusted node--free eight-user metropolitan quantum communication network},
  author={Joshi, Siddarth Koduru and Aktas, Djeylan and Wengerowsky, S{\"o}ren and Lon{\v{c}}ari{\'c}, Martin and Neumann, Sebastian Philipp and Liu, Bo and Scheidl, Thomas and Lorenzo, Guillermo Curr{\'a}s and Samec, {\v{Z}}eljko and Kling, Laurent and others},
  journal={Science advances},
  volume={6},
  number={36},
  pages={eaba0959},
  year={2020},
  publisher={American Association for the Advancement of Science}
}

@article{collett1984squeezing,
  title={Squeezing of intracavity and traveling-wave light fields produced in parametric amplification},
  author={Collett, MJ and Gardiner, CW},
  journal={Physical Review A},
  volume={30},
  number={3},
  pages={1386},
  year={1984},
  publisher={APS}
}

@article{alshowkan2021reconfigurable,
  title={Reconfigurable quantum local area network over deployed fiber},
  author={Alshowkan, Muneer and Williams, Brian P and Evans, Philip G and Rao, Nageswara SV and Simmerman, Emma M and Lu, Hsuan-Hao and Lingaraju, Navin B and Weiner, Andrew M and Marvinney, Claire E and Pai, Yun-Yi and others},
  journal={PRX Quantum},
  volume={2},
  number={4},
  pages={040304},
  year={2021},
  publisher={APS}
}

@article{lingaraju2021adaptive,
  title={Adaptive bandwidth management for entanglement distribution in quantum networks},
  author={Lingaraju, Navin B and Lu, Hsuan-Hao and Seshadri, Suparna and Leaird, Daniel E and Weiner, Andrew M and Lukens, Joseph M},
  journal={Optica},
  volume={8},
  number={3},
  pages={329--332},
  year={2021},
  publisher={Optica Publishing Group}
}

@article{qi2025multiuser,
  title={Multiuser quantum communication network via time-bin-entanglement-based frequency conversion and Bell-state measurement},
  author={Qi, Zhantong and Yang, Yilin and Wu, Chennan and Liao, Zixuan and Tang, Bo and Lei, Jiani and Li, Yuanhua and Lin, Jia and Zheng, Yuanlin and Chen, Xianfeng},
  journal={Physical Review A},
  volume={111},
  number={5},
  pages={052609},
  year={2025},
  publisher={APS}
}

@article{strobel2024quantum,
  title={Quantum Teleportation with Telecom Photons from Remote Quantum Emitters},
  author={Strobel, Tim and Vyvlecka, Michal and Neureuther, Ilenia and Bauer, Tobias and Sch{\"a}fer, Marlon and Kazmaier, Stefan and Sharma, Nand Lal and Joos, Raphael and Weber, Jonas H and Nawrath, Cornelius and others},
  journal={arXiv preprint arXiv:2411.12904},
  year={2024}
}

@article{luo2025entangling,
  title={Entangling quantum memories over 420 km in fiber},
  author={Luo, Xi-Yu and Wang, Chao-Yang and Zheng, Ming-Yang and Wang, Bin and Liu, Jian-Long and Gao, Bo-Feng and Li, Jun and Yan, Zi and Ke, Qiao-Mu and Teng, Da and others},
  journal={arXiv preprint arXiv:2504.05660},
  year={2025}
}

@article{qi202115,
  title={A 15-user quantum secure direct communication network},
  author={Qi, Zhantong and Li, Yuanhua and Huang, Yiwen and Feng, Juan and Zheng, Yuanlin and Chen, Xianfeng},
  journal={Light: Science \& Applications},
  volume={10},
  number={1},
  pages={183},
  year={2021},
  publisher={Nature Publishing Group UK London}
}

@article{stolk2024metropolitan,
  title={Metropolitan-scale heralded entanglement of solid-state qubits},
  author={Stolk, Arian J and van der Enden, Kian L and Slater, Marie-Christine and te Raa-Derckx, Ingmar and Botma, Pieter and Van Rantwijk, Joris and Biemond, JJ Benjamin and Hagen, Ronald AJ and Herfst, Rodolf W and Koek, Wouter D and others},
  journal={Science advances},
  volume={10},
  number={44},
  pages={eadp6442},
  year={2024},
  publisher={American Association for the Advancement of Science}
}

@article{ji2025global,
  title={A global quantum network with ground-based single-atom memories in optical cavities and satellite links},
  author={Ji, Jia-Wei and Sunami, Shinichi and Kikura, Seigo and Goban, Akihisa and Simon, Christoph},
  journal={arXiv preprint arXiv:2507.02333},
  year={2025}
}

@article{hu2025site,
  title={Site-selective cavity readout and classical error correction of a 5-bit atomic register},
  author={Hu, Beili and Sinclair, Josiah and Bytyqi, Edita and Chong, Michelle and Rudelis, Alyssa and Ramette, Joshua and Vendeiro, Zachary and Vuleti{\'c}, Vladan},
  journal={Physical Review Letters},
  volume={134},
  number={12},
  pages={120801},
  year={2025},
  publisher={APS}
}

@article{kikura2025passive,
  title={Passive Quantum Interconnects: High-Fidelity Quantum Networking at Higher Rates and Less Overhead},
  author={Kikura, Seigo and Tanji, Kazufumi and Goban, Akihisa and Sunami, Shinichi},
  journal={arXiv preprint arXiv:2507.01229},
  year={2025}
}

@article{duan2001long,
  title={Long-distance quantum communication with atomic ensembles and linear optics},
  author={Duan, L-M and Lukin, Mikhail D and Cirac, J Ignacio and Zoller, Peter},
  journal={Nature},
  volume={414},
  number={6862},
  pages={413--418},
  year={2001},
  publisher={Nature Publishing Group UK London}
}

\end{document}